 \documentclass[nohyper,12pt,letterpaper]{JHEP3}
 
 \def \eqn#1#2{\begin{equation}#2\label{#1}\end{equation}}

\title{Baryogenesis, Dark Matter and the Pentagon}

\author{T.\,Banks\\
Department of Physics and SCIPP\\
 University of California, Santa Cruz, CA 95064\\
E-mail: \email{banks@scipp.ucsc.edu}\\
{\it and}\\
Department of Physics and NHETC, Rutgers University\\
Piscataway, NJ 08540}

\author{S.\,Echols\\
Department of Physics and SCIPP\\
University of California, Santa Cruz, CA 95064\\
E-mail: \email{sechols@physics.ucsc.edu}}

\author{J.L.\,Jones\\
Department of Physics and SCIPP\\
University of California, Santa Cruz, CA 95064\\
E-mail: \email{jeff@physics.ucsc.edu}}

\abstract{We present a new mechanism for baryogenesis, which links
the baryon asymmetry of the universe to the dark matter density. The
mechanism arises naturally in the Pentagon model of TeV scale
physics.  In that context, it forces a re-evaluation of some of the
assumptions of the model, and we detail the changes that are
required in order to fit observations.}

 \received{} \accepted{}
\preprint{hep-ph/{0608104}\\\\RUNHETC-06-20 \\SCIPP-06-10}
\begin{document}

\section{\bf Introduction}

With a few exceptions\cite{exceptions} most models of the early
universe make no attempt to connect the observed baryon and dark
matter densities.   Dark matter is usually assumed to consist of
neutralinos or axions, and there is no connection between the
properties of these particles, and the baryon asymmetry.   In this
paper we will introduce a new class of models in which the solutions
to these two problems are directly related.   These models are
motivated by the Pentagon model of TeV scale physics\cite{remodel},
where the mechanism we will discuss arises naturally.   Indeed, it
is forced on us if we insist on finding a dark matter candidate
within the low energy model itself.

The basic new structure in this class of models is an approximate
symmetry, with current $J_{PB}^{\mu}$ .   This symmetry is
explicitly broken and an asymmetry $\epsilon$ for $J_{PB}^0$ is
generated in the very early universe.   The leading symmetry
violating operator at low energy has dimension $D$ and comes from
physics at a scale $M_b$.    The symmetry is also spontaneously
broken at a scale $f$.    Finally, there is a coupling in the low
energy effective Lagrangian of the form $${{1\over \Lambda^2}}
J_{PB}^{\mu} B_{\mu} ,$$ where $B_{\mu}$ is the ordinary baryon
number current.

The asymmetry in $J_{PB}$ then acts as an effective chemical
potential for ordinary baryon number.   Electroweak sphaleron
processes will then generate a baryon asymmetry (and a compensating
lepton asymmetry).    This is a form of spontaneous baryogenesis, a
mechanism invented by Cohen and Kaplan\cite{ck}.

At a lower energy scale, the effects of spontaneous and explicit
violation of $J_{PB}$ set in.   These convert the energy stored
initially in the asymmetry into a condensate of pseudo-Goldstone
bosons, which can be the dark matter.   We can use the parameters of
the model to fit both the baryon asymmetry, and dark matter density,
in a variety of ways.  However, in the version of the model that
arises from the Pentagon, all parameters but $\epsilon$ and $M_b$
are determined in terms of other experimental quantities.   We find
that, given the assumption in \cite{remodel} that $M_b$ is at least
as large as the neutrino seesaw scale ($5 \times 10^{14}$ GeV)the
baryon asymmetry is too large, or the dark matter density too small.
We describe modifications of the model which remove this problem in
section 4.

\section{Pseudo-goldstone dark matter with an asymmetry}

The idea of pseudo-Nambu-Goldstone bosons as dark matter is familiar
from axion models.   The general idea is that, after inflation, the
pseudo Goldstone field has a fixed value and small velocity.  Its
motion is friction dominated until the Hubble parameter falls to the
axion mass, at which time it begins to oscillate and behaves like
cold dark matter.  For appropriate values of the mass of the field
and its decay constant, a PNGB can reproduce the observed dark
matter density.

Here we propose an alternative initial condition for PNGB dark
matter.   The symmetry associated with the PNGB is explicitly
broken.   We assume that this breaking is larger during inflation
than it is at low energies.   A large value for the inflaton field,
typical in slow roll inflation models, can enhance operators that
are highly irrelevant at low energies.   We assume that this
produces an asymmetry $\epsilon$ at the end of inflation, and that
there is not much entropy production between the time this asymmetry
is generated, and the present.   Note that the Affleck-Dine
mechanism can easily generate large values of
$\epsilon$\cite{addrt}.

We note parenthetically that in string theory most PNGB's arise as
duals of fundamental anti-symmetric tensor gauge fields.   From this
point of view, the assumption of an asymmetry is equivalent to
assuming an expectation value for the magnetic field strength
$H_{ijk}$ and has been recently studied by Ibanez\cite{ibanez}. Our
work will be primarily devoted to PNGBs of accidental symmetries of
low energy gauge fields, but for completeness we will also study
what happens to a QCD axion with asymmetric initial conditions.

The non-zero value of $\epsilon$ and our assumption of isentropic
cosmological expansion allows us to write an equation for the time
dependence of the density $J_{PB}^0$

\eqn{dens}{J_{PB}^0 = \epsilon g T^3 ,} where $g$ is the effective
number of massless degrees of freedom.  This equation is valid from
the moment of asymmetry generation, until some time at which low
energy $J_{PB}^0$ violation becomes important, or a large amount of
entropy is dumped into the universe.  Note that it is valid both
above and below the scale of {\it spontaneous} breaking of the
approximate symmetry.   Below that scale, which we denote by $f$, we
have \eqn{spontdens}{J_{PB}^0 = f \partial_0 b ,} where $b$ is the
PNGB field. The primary focus of this paper is on models in which
$f$ is dynamically generated by strong gauge dynamics.

Assume that the explicit breaking of the symmetry after inflation
comes primarily from an operator of dimension $D$, and has its
origin in physics which decouples at the scale $M_b$.   Below the
scale $f$ this will give rise to an effective potential for $b$
\eqn{effpot}{{{\Lambda_L^D}\over {M_b^{D - 4}}} V(b/f).} For a QCD
axion we would have $D = 4$ and $\Lambda_L = \Lambda_{QCD}$. The
cross over temperature, at which this potential begins to affect the
evolution of the $b$ field is given by \eqn{t*}{{ (\epsilon g (T*)^3
)^2 \over 2} = {{f^{2}\Lambda_L^D}\over { M_b^{D - 4}}}, } or
\eqn{t*2}{\epsilon g (T*)^3 = \sqrt{2} f M_b^2 {\Lambda_L \over M_b}^{D/2}} At
this temperature, the ratio of energy densities of dark matter to
radiation is: \eqn{densrat}{\Big[{{\rho_b}\over
{\rho_{\gamma}}}\Big]_{T*} = {{(\epsilon g (T*)^3 )^2} \over {f^2 g
(T*)^4 }}.} Below this temperature, the $b$ field begins to
oscillate, and its energy scales like that of cold dark matter. The
dark matter to radiation ratio at any lower temperature is
\eqn{densrat2}{{{\rho_b}\over {\rho_{\gamma}}} = {{(\epsilon g
(T*)^3 )^2} \over {f^2 g (T*)^4 }}{{T*}\over T}.} An asymmetric PNGB
can be a good dark matter candidate if this ratio becomes one at
matter radiation equality, $T_{eq} \sim 1 {\rm eV}$. Using the
equation for $T*$, this requires \eqn{mbbnd}{M_b^{{D\over 2} - 2} =
{\epsilon\over f T_{eq}} \Lambda_L^{D/2}.}

\section{Spontaneous baryogenesis}

Suppose in addition that there is a coupling \eqn{currcoup}{{1\over
\Lambda^2 } J_{PB}^{\mu} B_{\mu} ,} between the current of the PNGB
and the ordinary baryon number current.   Assume for the moment that
$T* < T_{sh}$, where $T_{sh}$ is the scale below which the
electroweak baryon violating process shuts off, with $T_{sh} \sim
100$ GeV. Then, in the regime where electroweak baryon violation is
in equilibrium, this coupling has the form \eqn{chempot}{\epsilon g
{{T^3}\over \Lambda^2} B_0 \equiv \tilde{\mu} B_0 , } which is a
time dependent chemical potential for baryon number. The relative
rate of change of the chemical potential is the expansion rate of
the universe, much slower than sphaleron processes.   Thus, the
combination of $\tilde{\mu} $ and the sphaleron process put us in
the regime of equilibrium spontaneous baryogenesis, as defined by
Cohen and Kaplan\cite{ck}. Other baryon violating processes could
be hypothesized, at a variety of energy scales.  Their contribution
to the total baryon asymmetry would add to the one we compute here.
Without fine tuning, it is unlikely that the contributions of other
processes could cancel the electroweak sphaleron induced asymmetry.
It is straightforward to compute the induced baryon asymmetry (see
Appendix). We will also assume that any chemical potential for
lepton number (recall that $B - L$ is conserved by sphaleron
processes) is much smaller than that for baryon number, and that $B
- L$ asymmetries generated in the early universe are small compared
to the baryon asymmetry generated by our mechanism. We find
\eqn{barasymm}{\epsilon_B = {<B> \over {g V T^3}} =
{1\over2}\epsilon ({{T_{sh}}\over\Lambda})^2
 = 6\times 10^{-10} .} The last equality is the constraint from the observed
 baryon asymmetry.

 Note that if $T*$ is substantially larger than
 $T_{sh}$ then no asymmetry is generated because the chemical
 potential is turned off before baryon number violating processes go
 out of equilibrium.   The asymmetry will equilibrate to whatever
 value of the $B - L$ asymmetry was generated by early universe
 processes like leptogenesis.

Let us plug this value of $\epsilon$ into the formula we got by
insisting that the dark matter density comes out right. then we get
\eqn{flambdasquared}{M_b^{{D\over 2} - 2} = 1.2 \times 10^{-9}
{\Lambda^2 \over {f\ T_{sh}^2  T_{eq}}} \Lambda_L^{D/2} .} The
condition that $T* < T_{sh} $ translates into the inequality
 \eqn{t*bnd}{\Lambda_L^D < g M_b^{D-4} T_{sh}^3 T_{eq},}
or \eqn{t*bnd2}{\left({M_b\over \Lambda_L}\right)^{D/4} > {M_b \over g^{1/4} 0.18GeV} }
while the plausible constraint that $\epsilon < 10^3$ is
 \eqn{epsbnd2}{1.2 ({\Lambda \over T_{sh}})^2 < 10^{12}.}
For a QCD axion we expect $\Lambda > f$, where a strictly greater
than sign is used because the coupling between the axion field and
the baryon number current violates CP and might be suppressed by
more than dimensional analysis.   The conventional lower bounds on
the axion decay constant, from red giants and supernovae, then rule
out this kind of asymmetric axion scenario.  Note also that such a
scenario would have required a value of $\epsilon$ which is probably
too large to be generated by the Affleck-Dine mechanism.

Our analysis was based on the assumption that the pseudo-Goldstone
boson was the correct description of physics at the scale where the
primordial $J_{PB}^0 $ asymmetry is wiped out by processes which
violate this symmetry.   The temperature $T*$ where this occurs must
thus be smaller than $f$.   Scenarios where this inequality is not
satisfied are more complicated to analyze.   Some of them could give
rise to acceptable cosmologies.   However, both the QCD axion
models, and the Pentagon model, satisfy $T* < f$ so we will not
attempt to analyze this possibility any further.

\section{Spontaneous baryogenesis and dark matter in the Pentagon
model}

In the Pentagon model, all of the previous ingredients are present,
and most of the parameters are related.   There is a spontaneously
broken accidental symmetry, penta-baryon number.  It has $f = y$ TeV
and $ \Lambda_5 = x $ TeV, where $x$ and $y$ are of order $ 1 - 10 $
and most probably $y > x$ Using various different estimates of
standard model superpartner masses that have appeared in the
literature, we find $x$ running between $1.5$ and $7$. There is also
a current current coupling to baryon number with $ \Lambda_5 =
\alpha_3 \Lambda $, where the strong coupling is evaluated at the
TeV scale and is $\sim .1$.

The lowest dimension penta-baryon number violating operators, which
preserve gauge invariance, SUSY, and the fundamental discrete $R$
symmetry of the model are \eqn{d7op}{ \int d^2 \theta\ S P^5 ,} and
\eqn{d7op2}{ \int d^2 \theta\ S\tilde{P}^5 .}  These have dimension
$D = 7$\footnote{In \cite{remodel} TB forgot the R symmetry
constraint, and used the $D = 6$ operators without the singlet
$S$.}.   In \cite{remodel} TB also imposed the natural constraint
that all allowed irrelevant operators were suppressed by powers of
$M_U \sim 10^{15}$ GeV, which is the scale that appears in the
neutrino seesaw term.   Let us let the scale in the dimension $7$
operator be a free parameter, $M_b$.

The requirement that we get the right baryon asymmetry is
\eqn{baras}{\epsilon_B = {1\over 2} \epsilon ({{\alpha_3
T_{sh}}\over \Lambda_5})^2 ,} or \eqn{baras2}{\epsilon = 1.2 x^2
\times 10^{-5}.}  The equation determining $T*$ is
\eqn{t*}{{({\epsilon g (T*)^3})^2 \over 2 f^2} = {\Lambda_5^7 \over
M_b^3} ,} or \eqn{t*2}{\epsilon g (T*)^3 = \sqrt{2} f
{{\Lambda_5}^{7/2} \over M_b^{3/2}} .}  According to WMAP, the
temperature of matter radiation equality is about $1$ eV.   Thus,
the condition that the penton is dark matter is \eqn{eq}{1 = {\rho_b
\over \rho_{\gamma}} = {{(\epsilon g (T*)^3 )^2}\over {f^2 g
(T*)^4}}{T* \over T_{eq}}.}  If we insert the values for $\epsilon$
and $T_{eq}$ into this equation, we get the constraint \eqn{mb}{M_b
= ({x \over 2 y})^{2/3} x^3 \times 10^8 {\rm GeV}.}  The
prefactor is a number slightly less than $1$, so this ranges between
about $3 \times 10^8 \rightarrow 3 \times 10^{10}$ GeV as $x$ ranges
over the values allowed by the various estimates of superpartner
masses.  We also note that over the whole range of $x$, the
inequality $T* < T_{sh}$ is satisfied, as long as $y/x$ is not too
large.

The low scale $M_b$ raises the specter of unacceptably fast proton
decay. There is a dimension $6$ operator of the schematic form
$${1\over M^2} \int d^2\theta Q^3 L S ,$$ which is invariant under
all of the symmetries of the Pentagon model.   With $M = M_b$ this
would lead to disaster.    Thus, if we want to use the penton to
generate the dark matter in the universe, we must construct a theory
at the scale $M_b$ which explains the absence of operators of
dimension $6$, which could contribute to proton decay.

The alternative is to abandon the penton field $b$ as the origin of
dark matter, which is to say that $\epsilon$ is presumed to be
small.   We could retain a non-zero value of $\epsilon$ as the
mechanism for baryogenesis, though this would be no more attractive
than a host of other options.    Dark matter would have to come from
somewhere outside the Pentagon.   A possible candidate is an axion
dual to an antisymmetric tensor field.   This could also solve the
strong CP problem and be compatible with estimates from string
theory\cite{bdetc} .

Yet a third alternative, is to raise the values of $\Lambda_5$ and
$m_{ISS}$, probably abandoning the hypothetical connection of the
model to Cosmological SUSY Breaking.   A value of $\Lambda_5 \sim
3\times 10^5$ GeV and $M_b = 10^{15}$ GeV seems to be compatible
with both the observed dark matter density and baryon asymmetry. The
value of $\epsilon$ is a bit less than $1$ and $T*$ is well below
the sphaleron mass. However, there are a number of phenomenological
problems with this suggestion.  The rough estimates for sparticle
masses and the electroweak scale in the Pentagon model are
\eqn{gaugino}{ m_{1/2}^{(i)} \sim {50\over 3} g_S {{\alpha_i}\over
A\pi} m_{ISS} , }

\eqn{mer}{ m_{\tilde{e}_R} \sim \sqrt{50\over 3} {{\alpha_1}\over
B\pi} m_{ISS},} \eqn{ewscl}
 {H_u \sim 240
{\rm GeV} \sim {g_S \over 6} \Lambda_5 . } \eqn{msq}{m_{\tilde{q}}
\sim \sqrt{50\over 3} {{\alpha_3}\over B\pi} m_{ISS}.} $m_{ISS}$ is
the mass term which creates a meta-stable SUSY violating vacuum in
the Pentagon model.  The value of $A$ runs between $1$ and $8$,
while that of $B$ is between $1$ and $4$.  $g_S$ is a Yukawa
coupling between a next-to-minimal SSM singlet, and the Pentagon
fields. Other squark and slepton masses are larger than that of the
right handed selectron by factors of ${{\alpha_{2,3}}\over
{\alpha_1}}$.

 The factor of $1/6$ in the equation for the electroweak
scale represents a pious hope that the original Pentagon model with
$\Lambda_5 \sim 1.5$ TeV does not suffer from a {\it little
hierarchy problem}.   That is, the dimensional analysis estimate of
the electroweak scale has $1/6$ replaced by $1$, and we have to hope
that dynamical calculations in the strongly coupled Pentagon model
provide the factor of $6$.

If we postulate $\Lambda_5 \sim 3 \times 10^5$ GeV, dynamical
suppression is no longer plausible.  We can get the correct
electroweak scale by choosing $g_S$ small, about $5\times 10^{-3}$,
but this implies small gaugino masses.  The ratio between squark and
wino masses is

\eqn{ratio}{{m_{\tilde{q}}\over m_{\tilde{w}}} \sim {A\over 2B}
{{\alpha_3} \over \alpha_2} 10^2 .}  Using the experimental lower
bound on the wino mass we get squark masses that are a few times
$10$ TeV .   The model then has a hierarchy problem, and radiative
corrections to the Higgs mass are substantially larger than the
values indicated by precision electroweak fits.

Even a value $M_b = 10^{15}$ GeV is not enough to protect us from
proton decay.   The unified values of gauge couplings are quite
large in the Pentagon model, so even dimension six proton decay
operators must be suppressed by $10^{16}$ GeV or so.   We conclude
that raising the scale $\Lambda_5$ does not seem to be a promising
avenue for making a working model of penton dark matter and
baryogenesis.

\section{Conclusions}

The Pentagon model suggests a novel form of spontaneous
baryogenesis, which can tie together the dark matter density and the
baryon asymmetry.   While the general idea works quite well, it does
not work in the Pentagon model, unless we contemplate a scale of $
10^8  \rightarrow 10^{10} $ GeV for the leading irrelevant operator
that violates penta-baryon number.   It remains to be seen whether
one can invent a high energy extension of the model to generate this
operator without generating standard model operators already ruled
out by experiment.

If we stick with $10^{15}$ GeV as the scale for all irrelevant
corrections to the Lagrangian, then we must abandon the penton
theory of dark matter.   The penton is a light PNGB which might be
produced in the laboratory, but its cosmological abundance is
negligible.   It could still participate in the generation of the
baryon asymmetry we observe.

The most plausible candidate for dark matter compatible with the
Pentagon would then be a QCD axion.   This would also solve the
strong CP problem.

\section{Acknowledgments}

The authors would like to thank M.Dine for discussions about this
paper, and for suggesting that one could make the model work by
invoking a new scale of high energy penta-baryon violation. They
also want to thank the referee for pointing out a number of errors
in the original version of the manuscript. Their work was supported
in part by the Department of Energy under grant DE-FG03-92ER40689.


\section{Appendix }

In this appendix we go through the calculation of the baryon number
expectation, in the presence of a chemical potential in thermal
equilibrium.  We begin by ignoring leptons, which would play a role
in determining the equilibrium baryon number generated through
processes such as the electroweak sphaleron where the quantity $B-L$
is conserved.  After computing $\langle B\rangle$ first by ignoring
this constraint, we will then modify our calculation to include it.
This is mostly for pedagogical purposes, but the naive calculation
turns out to give the right order of magnitude as long as the lepton
chemical potential is of equal or smaller magnitude than the baryon
chemical potential.

Statistically, the equilibrium value of B can be calculated from the
partition function:
$$
\langle B\rangle = {\partial \over \partial \mu} \ln Z
$$
The $\mu$ being used here is dimensionless.  In terms of the usual
chemical potential $\tilde{\mu}$, which has dimensions of energy,
$\mu = {\tilde{\mu}\over T} = \tilde{\mu}\beta$.

The partition function for the baryons including the chemical
potential is

$$ Z = Tr(e^{-\beta H+\mu \int d^4x q_i^{\dagger}q_i}) $$
where
$$ H = \sum_{i,k} |k| [ a_k^{(i)\dagger}a_k^{(i)} + b_k^{(i)\dagger}b_k^{(i)} ] $$

$$ \int d^4x \thinspace q^{\dagger}q = {1 \over 3} \sum_k [ a_k^\dagger a_k - b_k^\dagger b_k ] $$

This can be re-expressed as
$$
Z = Tr[ \exp(-\sum_{i,k} ( \beta |k| - {\mu \over 3})
a_k^{(i)\dagger}a_k^{(i)}) \exp(- \sum_{i,k} ( \beta |k| +{\mu \over
3})  b_k^{(i)\dagger}b_k^{(i)})]
$$

$$
= Tr[\prod_{i,k} \exp(- ( \beta |k| - {\mu \over 3})
a_k^{(i)\dagger}a_k^{(i)}) \exp(- ( \beta |k| + {\mu \over 3})
b_k^{(i)\dagger}b_k^{(i)})]
$$
The trace gives a sum over all possible states of the system; for
fermions, each state can only have occupation number equal to zero
or one, thus
$$
Z = \prod_{k,i} (1 + e^{- ( \beta |k| - {\mu \over 3})} ) (1 + e^{-
( \beta |k| + {\mu \over 3})})
$$
We are interested in the logarithm of Z, which allows us to express
the logarithm of the product as a sum of logarithms
$$
\ln{Z} = 6 \times 3 \times 2 \times \sum_k [\ln (1 + e^{- ( \beta
|k| + {\mu \over 3}}) + \ln (1 + e^{- ( \beta |k| - {\mu \over
3})})]
$$
The factor of $6\times 3\times 2$ comes from the sum over $i$ which
includes six flavors of quarks, three colors, and two spin states.
Taking $k$ to the continuum limit we have
$$
\sum_k \rightarrow V \int {d^3 k \over (2 \pi)^3}
$$
and taking the $\mu$ derivative gives
$$
\langle B\rangle = {\partial \over \partial \mu} \ln Z =  36 V \int
{d^3 k \over (2 \pi)^3} [{1 \over 3 (
e^{\beta(k+{\mu\over3\beta})})} - {1 \over 3 (
e^{\beta(k-{\mu\over3\beta})})}]
$$
Using spherical coordinates, the angular integral just gives a
factor of $4\pi$, which leaves an integral over the magnitude k:
$$ \langle B\rangle = {6V \over \pi^2}(I_1-I_2)  $$

where
$$ I_1 = \int_0^{\infty} dk {{k^2 \over e^{\beta(k-{\mu\over3\beta})}+1}}$$
$$ I_2 = \int_0^{\infty} dk {{k^2 \over e^{\beta(k+{\mu\over3\beta})}+1}}$$

By substituting $x = \beta(k-{\mu\over3\beta})$, $I_1$ becomes:
$$ I_1 = {1 \over \beta^3} \int_{-{\mu\over3}}^{\infty} dx {(x+{\mu\over3})^2 \over e^x+1} = {1 \over \beta^3} \int_0^{\infty} dx {(x+{\mu\over3})^2 \over e^x+1} + {1 \over \beta^3} \int_{-{\mu\over3}}^0 dx {(x+{\mu\over3})^2 \over e^x+1} $$

and by substituting $x = \beta(k+{\mu\over3\beta})$, $I_2$ becomes:
$$ I_2 = {1 \over \beta^3} \int_{\mu\over3}^{\infty} dx {(x-{\mu\over3})^2 \over e^x+1} = {1 \over \beta^3} \int_0^{\infty} dx {(x-{\mu\over3})^2 \over e^x+1} - {1 \over \beta^3} \int_0^{\mu\over3} dx {(x-{\mu\over3})^2 \over e^x+1} $$

The two can then be recombined into:

$$ I_1 - I_2 = {1 \over \beta^3} \Big[\int_0^{\infty} dx {4x ({\mu\over3})\over e^x+1} + \int_{-{\mu\over 3}}^0 dx {(x+{\mu\over3})^2 \over e^x+1} + \int_{\mu \over3}^0 dx {(x-{\mu\over3})^2 \over e^x + 1}\Big]$$

$$ I_1 - I_2 = {1 \over \beta^3} \Big[ ({4\mu \over 3})({\pi^2 \over 12}) + \int_0^{\mu \over3} dx \big\{ {(x-{\mu\over3})^2 \over e^{-x}+1} + {(x-{\mu\over3})^2 \over e^{x}+1} \big\} \Big] $$

Miraculously, the exponentials simply add up to 1, leaving an
integral of a polynomial:

$$ = {1 \over \beta^3} \Big[ {\pi^2 \mu \over 9} + \int_0^{\mu \over 3} dx (x-{\mu \over3})^2 \Big] $$

$$ = {1 \over \beta^3} [ {\pi^2 \over 3} ({\mu \over 3}) + {1 \over 3} ({\mu \over3})^3 ] $$

$$ \langle B\rangle = {2V\over 3\beta^3} (\mu + {1\over27\pi^2}\mu^3) $$

The above discussion, however, was naive in that we have ignored the
role of leptons, ie this calculation does not conserve $B-L$.
Therefore, we must modify the original partition function by
including a chemical potential for leptons similar to that for
baryons, and impose this constraint by including a Kronecker delta
function:

$$
Z = Tr[e^{-\beta H}e^{\mu_B \int d^4x q_i^{\dagger}q_i}e^{\mu_L \int
d^4x l_i^{\dagger}l_i} \delta_{BL}]$$

where now

$$ H = \sum_{i,k} |k| [ a_k^{(i)\dagger}a_k^{(i)} + b_k^{(i)\dagger}b_k^{(i)} + c_k^{(i)\dagger}c_k^{(i)} + d_k^{(i)\dagger}d_k^{(i)}  ] $$

with the Baryon and Lepton operators expressed as

$$ \int d^4x \thinspace q^{\dagger}q = {1 \over 3} \sum_k [ a_k^\dagger a_k - b_k^\dagger b_k ], $$

$$ \int d^4x \thinspace l^{\dagger}l = \sum_k [ c_k^\dagger c_k - d_k^\dagger d_k ]. $$

The delta function can be represented in integral form as

$$ \delta_{BL} = {1\over 2\pi} \int_{-\pi}^{\pi} d\alpha e^{i \alpha(B-L)}$$
(We can choose any range covering a full period of $2\pi$ for the
integration limits, but as we will find later, the range from $-\pi$
to $\pi$ turns out to be particularly convenient.)

$$B-L = \sum_{k} ({1 \over 3} a_k^{(i)\dagger}a_k^{(i)} - {1 \over 3} b_k^{(i)\dagger}b_k^{(i)} - c_k^{(i)\dagger}c_k^{(i)} + d_k^{(i)\dagger}d_k^{(i)})
$$

It follows that

$$
Z = Tr \Big[{1 \over 2 \pi} \int_{-\pi}^{\pi} d\alpha (\prod_{i,k}
\exp( (- \beta |k| + {\mu_B \over 3} + {i \alpha \over 3})
a_k^{(i)\dagger}a_k^{(i)})) (\prod_{i,k} \exp( (- \beta |k| - {\mu_B
\over 3} -{i \alpha \over 3} )  b_k^{(i)\dagger}b_k^{(i)}))$$
$$\space \space \space \times (\prod_{i,k} \exp( (- \beta |k| + {\mu_L} - {i \alpha})  c_k^{(i)\dagger}c_k^{(i)})) (\prod_{i,k} \exp( (- \beta |k| - {\mu_L} +{i \alpha} )  d_k^{(i)\dagger}d_k^{(i)})) \Big].
$$

As before, the trace gives a sum over all possible states of the
system (zero or one for fermions), thus:

$$
Z = {1 \over 2 \pi} \int_{-\pi}^{\pi} d\alpha (\prod_{k,i} 1 + e^{
(- \beta |k| + {\mu_B \over 3} + {i \alpha \over 3} )} )
(\prod_{k,i} 1 + e^{ (- \beta |k| - {\mu_B \over 3} - {i \alpha
\over 3} )})$$
$$\space \space \space \times (\prod_{k,i} 1 + e^{ (- \beta |k| + {\mu_L } - {i \alpha} )} ) (\prod_{k,i} 1 + e^{ (- \beta |k| - {\mu_L } + {i \alpha } )})
$$

Unlike the case for baryons alone, we cannot exploit the fact that
we wish to find the logarithm of Z to convert this product into a
sum of logarithms because of the $\alpha$ integral.  However, we can
write

$$
Z = {1 \over 2 \pi} \int_{-\pi}^{\pi} d\alpha \exp \Big[  V \int
{d^3 k \over (2 \pi)^3} 36[ \ln (1 + e^{ (- \beta |k| + {\mu_B \over
3} + {i \alpha \over 3} )}) + \ln (1 + e^{ (- \beta |k| - {\mu_B
\over 3} - {i \alpha \over 3} )})]$$
$$ +  9[ \ln (1 + e^{ (- \beta |k| + {\mu_L } - {i \alpha } )}) + \ln (1 + e^{ (- \beta |k| - {\mu_L } + {i \alpha } )})] \Big]
$$

Again we have taken $k$ to the continuum limit.  As before the
baryons obtain a factor of 36 from the sum over $i$ (6 flavors, 3
colors, 2 spin states), the leptons similarly obtain a factor of 3
generations $\times$ 3 spin states (two for each charged lepton and
one each neutrino)  $=9$.

The integral in the exponent is the same integral we had in the
purely baryonic case, but with the real parameter $\mu / 3$ replaced
by a complex parameter $z$, where $z = (\mu_B +i\alpha)/3$ for the
baryonic terms, and $z = \mu_L-i\alpha$ for the leptonic terms.  The
integrand has poles along the imaginary axis of the $z$-plane at
Im$\{z\}=\pi+2\pi n$, $n$ integer, and they occur when $k = {\rm
Re}\{z\}$.  The analytic form of the integral we found for real $z$
above,

$$ {1 \over \beta^3} \big[{\pi^2 \over 6} z^2 + {1\over 12} z^4 + {\rm const} \big]$$
is valid as long as Im$\{z\}$ is restricted to the range
$(-\pi,\pi)$.  For Im$\{z\}$ outside of this range, the function
simply repeats with period $2\pi$, which happens because the
integrand was periodic in $\alpha$.  The closed form expression for
the full periodic function can be written by replacing $z$ with $\ln
e^z$ in the polynomial.  As long as we choose the range of
integration over $\alpha$ to be from $-\pi$ to $\pi$, the original
polynomial form above can be used.  The partition function is then:

$$Z = {1\over 2\pi} \int_{-\pi}^{\pi} d\alpha \exp[{1\over 2\pi^2\beta^3}f(\alpha,\mu_B,\mu_L)+{\rm const}]$$
where
$$ f(\alpha,\mu_B,\mu_L) = 36\big[{\pi^2\over 6} ((\mu_B+i\alpha)/3)^2 + {1\over 12} ((\mu_B + i\alpha)/3)^4 \big]
 + 9 \big[ {\pi^2\over 6} (\mu_L - i\alpha)^2 + {1\over12}(\mu_L - i\alpha)^4\big]$$

The constant term is independent of both $\mu_B$ and $\mu_L$.  Since
we are only interested in expectation values which do not depend on
the normalization of the partition function, we can ignore this
term.

The $\alpha$ integral can be performed using the saddle point
approximation.  Allowing $\alpha$ to be complex now, there is a
maximum in Re$\{f\}$ at Re$\{\alpha\} = 0$.  The saddle point occurs
at the point on the imaginary axis where there is a minimum in
Re$\{f\}$ along the imaginary direction in the $\alpha$-plane.  In
the text, we will find that $\mu_{L,B}$ must be small in order to
fit the observed baryon asymmetry.   Therefore, we will work to
lowest order in the chemical potentials, and avoid writing the
lengthy exact expression. \vfill\eject \noindent To first order in
$\mu_L$ and $\mu_B$, the saddle point is at:

$$\alpha_{\rm saddle} = {4i\over 13}\mu_B-{9i \over 13}\mu_L$$

The value of $f$ at $\alpha_{\rm saddle}$, to second order in
$\mu_B$ and $\mu_L$ is:
$$f(\alpha_{\rm saddle},\mu_B,\mu_L) = 9.11\mu_B\mu_L+4.55\mu_B^2+4.55\mu_L^2$$

$Z$ can be approximated well by evaluating the integrand at
$\alpha_{\rm saddle}$.  In the saddle point approximation, this
would usually need to be divided by the square root of the second
derivative of $f$ with respect to $\alpha$, but this just leads to
an additive term in $\ln Z$ which is independant of the volume of
spacetime.  As $V \rightarrow \infty$, this is negligible compared
to the term proportionate to $V$.

$$\ln Z \sim {V\over 2\pi^2\beta^3} f(\alpha_{\rm saddle},\mu_B,\mu_L) = 0.46 {V\over \beta^3}(\mu_B\mu_L + {1\over2} \mu_B^2 + {1\over2}\mu_L^2) + O(\mu^3_{B,L})$$

Therefore, the expectation values of baryon and lepton number are,
to first order in $\mu_B$ and $\mu_L$:
$$\langle B\rangle = \langle L\rangle = {\partial\ln Z\over \partial\mu_B} = {\partial\ln Z\over \partial\mu_L} \approx {V \over 2\beta^3}(\mu_B + \mu_L)$$
This calculation gives us the baryon asymmetry written in Equation
3.3, when we plug in the effective chemical potential in Equation
3.2.




  %




\end{document}